\documentclass[12pt]{article}
\pdfoutput=1
\usepackage{latexsym, graphicx,cite}
\usepackage{amsmath}
\usepackage{amssymb}
\usepackage{amsthm}
\allowdisplaybreaks

\usepackage[top = 1. in,bottom = 1. in, left = 1 in, right=1 in]{geometry}

\newcommand{\arXiv}[1]{\href{http://www.arXiv.org/abs/#1}{arXiv:#1}}
\usepackage[colorlinks=true, linkcolor=blue, bookmarks=true]{hyperref}

\makeatletter
\renewcommand\section{\@startsection {section}{1}{\z@}%
        {-3.5ex \@plus -1ex \@minus -.2ex}
         {2.3ex \@plus.2ex}%
         {\normalfont\large\bfseries}}
\renewcommand\subsection{\@startsection{subsection}{2}{\z@}%
          {-3.25ex\@plus -1ex \@minus -.2ex}%
          {1.5ex \@plus .2ex}%
          {\normalfont\bfseries}}
\makeatother

\newcommand{\beq}{\begin{equation}}
\newcommand{\eeq}{\end{equation}}
\newcommand{\ber}{\begin{array}}
\newcommand{\eer}{\end{array}}

\newcommand{\de}{\delta}

\newcommand{\eps}{\varepsilon}

\newcommand{\ena}{\end{eqnarray}}
\newcommand{\beqa}{\begin{eqnarray}}
\newcommand{\eeqa}{\end{eqnarray}}
\newcommand{\bea}{\begin{eqnarray}}
\newcommand{\eea}{\end{eqnarray}}

\theoremstyle{remark}

\renewcommand{\Re}{\operatorname{Re}}

\begin{document}
\begin{titlepage}
\begin{flushright}
\phantom{arXiv:yymm.nnnn}
\end{flushright}
\vspace{1cm}
\begin{center}
{\LARGE\bf A nonrelativistic limit for AdS perturbations}\\
\vskip 15mm
{\large Piotr Bizo\'n,$^a$ Oleg Evnin$^{b,c}$ and Filip Ficek$\hspace{0.2mm}^a$}
\vskip 7mm
{\em $^a$ Institute of Physics, Jagiellonian University, Krak\'ow, Poland}
\vskip 3mm
{\em $^b$ Department of Physics, Faculty of Science, Chulalongkorn University,
Bangkok, Thailand}
\vskip 3mm
{\em $^c$ Theoretische Natuurkunde, Vrije Universiteit Brussel and\\
The International Solvay Institutes, Brussels, Belgium}
\vskip 7mm
{\small\noindent {\tt bizon@th.if.uj.edu.pl, oleg.evnin@gmail.com, filip.ficek@doctoral.uj.edu.pl}}
\vskip 10mm
\end{center}
\vspace{1cm}
\begin{center}
{\bf ABSTRACT}\vspace{3mm}
\end{center}

The familiar $c\to \infty$ nonrelativistic limit converts the Klein-Gordon equation in Minkowski spacetime to the free Schr\"odinger equation, and the Einstein-massive-scalar system without a cosmological constant to the Schr\"odinger-Newton (SN) equation. In this paper, motivated by the problem of stability of Anti-de Sitter (AdS) spacetime, we examine how this limit is affected by the presence of a negative cosmological constant $\Lambda$. Assuming for consistency that the product $\Lambda c^2$ tends to a negative constant as $c\to\infty$, we show that the corresponding nonrelativistic limit is given by the SN system with an external harmonic potential which we call the Schr\"odinger-Newton-Hooke (SNH) system. We then derive the resonant approximation which captures the dynamics of small amplitude spherically symmetric solutions of the SNH system. This resonant system turns out to be much simpler than its general-relativistic version, which makes it amenable to analytic methods. Specifically, in four spatial dimensions, we show that the resonant system possesses a three-dimensional invariant subspace on which the dynamics is completely integrable and hence can be solved exactly. The evolution of the two-lowest-mode initial data (an extensively studied case for the original general-relativistic system), in particular, is described by this family of solutions.

\vfill

\end{titlepage}


\section{Introduction}

Numerical and perturbative studies of dynamics of small perturbations of the Anti-de Sitter (AdS) spacetime in \cite{BR} have led to a conjecture that the AdS spacetime is nonlinearly unstable, with black hole formation in the evolution of arbitrarily small generic perturbations. While significant further evidence has been accumulated for this conjecture in subsequent works, the regime of very small perturbations is not accessible numerically, which necessitates the development of exact or asymptotic analytic methods. The only rigorous result that we are aware of is the proof of instability of AdS within the spherically symmetric massless Einstein-Vlasov model with negative cosmological constant announced in \cite{M2} (see also \cite{M1}). However, the argument in \cite{M2} relies heavily on the fact that matter moves along null geodesics and it is not clear a priori whether a similar approach can be applied to other Einstein-matter or vacuum Einstein equations.

Another strategy, taken here, is based on the resonant approximation developed and applied in~\cite{FPU,CEV1,CEV2,BMR,islands,rev2} under a number of different names (two-time framework, renormalization flow, time-averaging). Within this approximation, the dynamics of small perturbations of AdS is described by an infinite dimensional cubic dynamical system for the normal mode amplitudes. In this representation, the magnitude of perturbations in the original system is completely scaled out, thereby giving access to the regime of arbitrarily small perturbations. Numerical analysis \cite{BMR} of the resonant approximation for the Einstein-scalar system in $4+1$ dimensions with negative cosmological constant provided further evidence that the instability of AdS is triggered by a turbulent transfer of energy to short wavelength modes, as conjectured in \cite{BR}.

Even though the resonant approximation provides an attractive window into extremely small amplitude regime of AdS perturbations, it is still forbiddingly complicated for analytic purposes. This is ultimately traced back to the complexity of nonlinearities in Einstein's equations. The coefficients of the resonant system are complicated expressions in terms of integrals of the AdS mode functions \cite{CEV1,CEV2}, and in special cases where these integrals have been evaluated \cite{islands}, the expressions are so long that they could not even be included in a published article. In such circumstances, it seems wise to look at simpler related systems. Most pragmatically, understanding general features of resonant systems may lead to qualitative results that will shed light on the dynamics of AdS perturbations irrespectively of the detailed algebraic expressions for the coefficients in the corresponding resonant system. More broadly, this point of view connects weakly nonlinear dynamics of asymptotically AdS spacetimes to many other interesting studies of weakly nonlinear energy transfer in spatially confined systems, often motivated by completely different physics.

Perhaps the most straightforward modification of the dynamics of asymptotically AdS spacetimes is turning off the gravitational backreaction and studying nonlinear matter fields on a fixed AdS background \cite{BKS,CF,BHP1,BEL,BHP2}. While this is a very different physical setting, the corresponding resonant approximation has an identical structure but with much simpler coefficients in the equations. Finite-time turbulent blow-ups are no longer seen but the dynamics is still characterized by a sequence of direct cascades of energy to shorter wavelength modes, followed by inverse cascades. Some of such resonant systems originating from nonlinear wave equations in AdS can be treated analytically with a number of explicit results \cite{CF,BHP1,BEL,BHP2}. (These systems show some similarities to the cubic Szeg\H o equation designed and studied in a series of works by G\'erard and Grellier starting from \cite{GG}.
Remarkably, the cubic Szeg\H o
\nopagebreak[0]
equation exhibits weak turbulence even though it is integrable.)

There is a deep similarity between nonlinear wave equations in AdS and the Gross-Pitaevskii equation describing Bose-Einstein condensates in harmonic traps (the latter system is nowadays routinely realized experimentally in terrestrial  laboratories). This similarity has been emphasized at a practical level in \cite{BMP}, and explained in \cite{BEL} through the emergence of the Gross-Pitaevskii equation as the nonrelativistic limit\footnote{At the level of the corresponding symmetry groups, rather than at the level of the equtions of motion, this relation has appeared in the literature a number of times \cite{BLL, DD, Aldrovandi:1998im, Gibbons:2003rv}.} of the cubic wave equation in AdS. Through this connection, a number of interesting parallels have been observed between the resonant systems for AdS wave equations and the Gross-Pitaevskii equations, and the corresponding exact analytic solutions in weakly nonlinear regimes \cite{GHT,GT,BBCE,GGT,BBCE2,Bia18}.

Given the above connections (interesting in both physical and mathematical terms) that emerge from taking a nonrelativistic limit of nonlinear wave equations in AdS, a question arises: What if one takes a similar nonrelativistic limit in the original AdS instability setup, with the full gravitational backreaction of the scalar field included? This is the main question we shall be addressing in this article. We recall that the corresponding nonrelativistic limit of the Einstein-massive-scalar system with zero cosmological constant (i.e., for asymptotically flat rather than asymptotically AdS spacetimes), resulting in the Schr\"odinger-Newton (SN) equation for a self-gravitating wavefunction, has been considered in various physical contexts,\footnote{Outside gravitational physics, the SN equation occurs (under the names of the Hartree, Schr\"odinger-Poisson or Choquard equation) as a mean-field approximation for many-body problems; see \cite{choq} for a review.} for example  in modelling boson stars \cite{Kaup, RB} or in attempts to envisage the wave function collapse as a gravitational phenomenon \cite{Penrose,MPT}. It has also been extensively studied in the mathematical literature (see, e.g., \cite{KLR, KMR}).

 In the presence of a negative cosmological constant $\Lambda$, an important observation is that consistency of the nonrelativistic limit requires the product $\Lambda c^2$ to approach a negative constant as $c\rightarrow \infty$. Under this assumption, we show that the nonrelativistic limit is given by the SN system with an external harmonic potential. Thus, the confinement of waves in asymptotically AdS spacetimes due to the gravitational potential
 translates in the nonrelativistic limit to the trapping by the harmonic potential. We call this system the Schr\"odinger-Newton-Hooke (SNH) equation following the tongue-in-cheek terminology for the Newton-Hooke groups in the literature on kinematic symmetries \cite{Aldrovandi:1998im, Gibbons:2003rv}.
 Although the SNH system has been studied in the literature (see, e.g., \cite{CMS, FL}), we believe
 it is the first time it appears in connection with a negative cosmological constant.

The paper is organized as follows. In section~2, we derive the SNH system as the nonrelativistic limit of the
Einstein-massive-scalar system with a negative cosmological constant and discuss its basic properties. We point out that in four spatial dimensions the SNH system enjoys a symmetry enhancement analogous to the symmetry enhancement for the Gross-Pitaevskii equation in two spatial dimensions \cite{Niederer,OFN}. In section~3, we construct the resonant system which approximates the small amplitude regime of the SNH equation. In section~4, we study this resonant system and prove that in four spatial dimensions it belongs to a large class of analytically tractable resonant systems developed in \cite{Bia18}. This automatically gives exact special analytic solutions, bounds on turbulent energy transfer for these special solutions, and an extra conserved quantity. We conclude with a discussion of the SNH system in higher dimensions.


\section{The SNH model}

We consider a massive self-gravitating scalar field $\phi$ in $d+1$ dimensions (for $d\geq 3$) in the presence of negative cosmological constant $\Lambda$. The system is governed by the Klein-Gordon equation
\begin{equation}\label{KG}
 g^{\mu\nu} \nabla_{\mu}\nabla_{\nu} \phi - \frac{m^2 c^2}{\hbar^2} \phi=0,
\end{equation}
and the
Einstein equations
\begin{equation}\label{EE}
 G_{\mu \nu} + \Lambda g_{\mu \nu} = \frac{8\pi G}{c^4}\, T_{\mu\nu}
\end{equation}
with the stress-energy tensor
\begin{equation}\label{T}
 T_{\mu\nu} = \frac{\hbar^2}{2m} \left[\partial_{\mu} \phi \partial_{\nu} \bar \phi + \partial_{\mu} \bar \phi \partial_{\nu} \phi - g_{\mu\nu} \left(\partial_{\lambda} \phi \partial^{\lambda} \bar \phi
 +\frac{m^2 c^2}{\hbar^2} |\phi|^2\right)\right]\,.
\end{equation}
To derive a nonrelativistic limit,\footnote{The nonrelativistic limit for the Einstein-scalar system in four spacetime dimensions without cosmological constant was examined in details in \cite{Giu12} under the assumption of spherical symmetry (see also \cite{FF} for the case with positive cosmological constant).} we assume the following weak-field ansatz for the leading order behavior (in powers of $1/c$) of the scalar field and the metric (hereafter $x\in \mathbb{R}^d$):
 \begin{equation}\label{psi}
  \phi(t,x)=e^{-\frac{i m c^2}{\hbar} t}\,u(t,x)\,,
 \end{equation}
 and
\begin{equation}\label{metric}
 ds^2 = -c^2 \left(1+\frac{2A(t,x)}{c^2}\right)\,dt^2+\left(1+\frac{2B(t,x)}{c^2}\right) \sum_{j=1}^d (dx^j)^2\,.
\end{equation}
Note that in the ansatz \eqref{metric} we have included the first post-Newtonian correction to the spatial part of the metric which at this order in $1/c$ expansion is isotropic; see \cite{MTW} for a textbook treatment of a similar nonrelativistic limit without a cosmological constant in $3+1$ dimensions.
We insert this ansatz into the field equations and determine the leading order terms in powers of $1/c$. A salient point here is that we must assume for consistency that $\Lambda \rightarrow 0$ as $c\rightarrow \infty$. More precisely, we require that\footnote{An explanation for the group-theoretical origin of \eqref{omega} in the context of taking nonrelativistic limits can be found in \cite{Gibbons:2003rv}. We are grateful to Gary Gibbons for elucidating this point to us.}
\begin{equation}\label{omega}
 \lim_{c\rightarrow \infty} \Lambda c^2 = -\frac{d(d-1)}{2}\,\omega^2,
\end{equation}
where $\omega$ is a constant. The Klein-Gordon equation \eqref{KG} gives for $c\rightarrow \infty$
 \begin{equation}\label{schr0}
  i \hbar \,\partial_t u = - \frac{\hbar^2}{2m} \, \Delta u + m A u\,,
 \end{equation}
 where $\Delta=\sum_{j=1}^d \partial_j^2$.
 \pagebreak

 \noindent The $tt$-component of the Einstein equations gives for $c\rightarrow \infty$
 \begin{equation}\label{Gtt}
  -(d-1) \Delta B +\frac{d(d-1)}{2} \omega^2 = 8\pi G m |u|^2,
 \end{equation}
 while the $jj$-component multiplied by $c^2$ gives
 \begin{equation}\label{Gjj}
  \Delta A -\partial_j^2 A + (d-2) (\Delta B -\partial_j^2 B)- \frac{d(d-1)}{2} \omega^2 = 0.
 \end{equation}
 Summing this up over $j$ and using \eqref{Gtt} we get
\begin{equation}\label{poisson0}
 \Delta A = \frac{8\pi G (d-2)}{d-1} |u|^2 +  \omega^2 d\,.
\end{equation}
Substituting $A=V+ \frac{1}{2}\omega^2 |x|^2$ into \eqref{schr0} and \eqref{poisson0}, we obtain
\begin{align}
i\hbar \, \partial_t u&=-\frac{\hbar^2}{2m}\Delta u+\frac{1}{2}m\omega^2 |x|^2 u+V u, \label{schr}\\
\Delta V&= \frac{8\pi G (d-2)}{d-1} |u|^2\,,\label{poisson}
\end{align}
which we shall call the Schr\"odinger-Newton-Hooke (SNH) system. The Poisson equation \eqref{poisson}
can be solved using the Green function for the Laplace operator
\begin{equation}\label{green}
 G(x)=-\frac{1}{(d-2) \Omega_d} \, \frac{1}{|x|^{d-2}},
\end{equation}
where $\Omega_d=2\pi^{\frac{d}{2}}/\Gamma(\frac{d}{2})$ is the volume of $\mathbb{S}^{d-1}$,
to yield
\begin{equation}\label{Vsol}
V(t,x) =-\frac{8\pi G m}{(d-1) \Omega_d} \, \int_{\mathbb{R}^d} \frac{|u(t,y|^2}{|x-y|^{d-2}}\, dy\,,
\end{equation}
which upon substitution into \eqref{schr} gives the Hartree equation with a harmonic potential
\begin{equation} \label{eqH}
i \hbar \, \partial_t u=-\frac{\hbar^2}{2m}\Delta u+\frac{1}{2} m \omega^2 |x|^2 u-\frac{8\pi G m}{(d-1) \Omega_d} \, \left(\int_{\mathbb{R}^d} \frac{|u(t,y)|^2}{|x-y|^{d-2}}\right) \, u\,.
\end{equation}
In this paper we restrict ourselves to spherically symmetric solutions. In this case, the convolution can be simplified using Newton's formula
\begin{equation}\label{newton}
 \int_{\mathbb{R}^d} \frac{f(|y|)}{|x-y|^{d-2}}\, dy=
 \int_{\mathbb{R}^d} \frac{f(|y|)}{\max\{|x|^{d-2},|y|^{d-2}\}}\, dy=\Omega_d \int_0^{\infty} \frac{f(s) s^{d-1}}{\max\{r^{d-2},s^{d-2}\}}\, ds\,.
\end{equation}
Setting $\hbar=m=\frac{8\pi G}{d-1}=1$ by a choice of units and using \eqref{newton}, at the end we arrive at the radial Hartree equation with a harmonic potential
\begin{equation} \label{eqSNH}
i \partial_t u=-\frac{1}{2} \Delta_r u +\frac{1}{2} \omega^2 r^2 u-\left(\int_0^\infty \frac{|u(t,s)|^2 s^{d-1}}{\max\{r^{d-2},s^{d-2}\}}\, ds\right) u\,,
\end{equation}
where $\Delta_r=\partial_r^2+\frac{d-1}{r} \partial_r$. We shall refer to \eqref{eqSNH} as the SNH equation.

The SNH equation preserves the mass and energy defined as
\begin{align}
\mathcal{N}[u]&= \int_0^{\infty} |u(t,r)|^2 r^{d-1} dr,\\
\mathcal{H}[u]&= \frac{1}{2} \int_0^{\infty} |\partial_r u(t,r)|^2 r^{d-1} dr
+\frac{1}{2}\omega^2 \int_0^{\infty} r^2 |u(t,r)|^2 r^{d-1} dr \nonumber \\&\qquad-\frac{1}{4} \int_0^{\infty} \left(\int_0^\infty \frac{|u(t,s)|^2 s^{d-1}}{\max\{r^{d-2},s^{d-2}\}}\, ds\right) |u(t,r)|^2 r^{d-1} dr\,.
\end{align}

We shall now display a connection between the SNH equation (\ref{eqSNH}) and the SN equation, which results from setting $\omega=0$ in (\ref{eqSNH}):
\begin{align}
i \partial_t u=-\frac{1}{2}\Delta_r u-\left(\int_0^\infty \frac{|u(t,s)|^2 s^{d-1}}{\max\{r^{d-2},s^{d-2}\}}\, ds\right) u\,.
\label{eqn:now}
\end{align}
This equation enjoys the scaling symmetry
\begin{align}
u(t,r)\mapsto u_{\lambda}(t,r):=\lambda^2 u(\lambda^2 t,\lambda r).
\label{eqn:scaling}
\end{align}
Under this scaling, the mass and energy corresponding to (\ref{eqn:now}) transform as follows
\begin{equation}\label{clas}
\mathcal{N}_0[u_{\lambda}]=\lambda^{4-d} \mathcal{N}_0[u],\qquad \mathcal{H}_0[u_{\lambda}]= \lambda^{6-d} \mathcal{H}_0[u],
\end{equation}
hence the SN equation is mass-critical for $d=4$ and energy-critical for $d=6$. Although the scaling symmetry
 is broken in \eqref{eqSNH} by the harmonic term, these critical dimensions demarcate different behaviors of solutions of the SNH equation as well. In particular, in the mass critical case, there is a direct relation between the SN and SNH equations. To see this, given a solution $u(t,r)$ of equation \eqref{eqn:now}, we define the lens transform $\L u$ as in \cite{Carles02}:
\begin{align}
\L u(t,r)= \frac{1}{\cos^{d/2} \omega t}\, u\left(\frac{\tan \omega t}{\omega},\frac{r}{\cos \omega t}\right)e^{-\frac{i}{2}\omega r^2 \tan \omega t}.
\label{eqn:lens1}
\end{align}
It can be verified by a direct computation that $\L u$ satisfies
\begin{align}
i\partial_t \L u=-\frac{1}{2}\Delta_r \L u+\frac{1}{2}\omega^2 r^2\, \L u-\cos^{d-4} (\omega t) \left(\int_0^\infty \frac{|\L u(t,s)|^2 s^{d-1}}{\max\{r^{d-2},s^{d-2}\}} ds\right) \L u,
\label{eqn:lens2}
\end{align}
which for $d=4$ simply reduces to (\ref{eqSNH}). Note that $\L u(0,r)=u(0,r)$, so the lens transform does not alter the initial data. Thus, solving \eqref{eqSNH} in $d=4$ is equivalent to solving (\ref{eqn:now}) for the same initial data, and then applying the lens transform to obtain the desired solution. This structure evidently implies that all symmetries of (\ref{eqn:now}), given by the Schr\"odinger group \cite{DL} in $d=4$ are conjugated to the analogous symmetries of (\ref{eqSNH}). Note, however, that if $u(t,r)$ is a global-in-time solution of \eqref{eqn:now}, then $\L u(t,r)$ is defined only for $|t|<\frac{\pi}{2\omega}$ and it is a delicate question what happens for later times.
The situation is directly parallel to the Gross-Pitaevskii equation, where both the lens transform \cite{Carles02, Tao09} and the underlying conformal symmetry \cite{Niederer,OFN} operate in $d=2$. Similarly, our subsequent study of weakly nonlinear dynamics will reveal further strong parallels between the SNH equation in $d=4$ and the Gross-Pitaevskii equation with an isotropic harmonic potential in $d=2$.

The lens transform (\ref{eqn:lens1}) is a nonrelativistic analog of the familiar coordinate and conformal transformations encountered in AdS spacetimes. Namely, Minkowski spacetime with the flat metric $ds^2=-dt^2+dz^2+dx^idx^i$, whose massive scalar perturbations are described in the nonrelativistic limit by the SN equation, is conformal to the Poincar\'e patch of AdS spacetime with the metric $ds^2=(-dt^2+dz^2+dx^idx^i)/z^2$. By a standard change of coordinates, the Poincar\'e patch is embedded into the global AdS spacetime, whose massive scalar perturbations are described in the nonrelativistic limit by the SNH equation. This map between a system with curvature (global AdS) and a system without curvature (Minkowski) is directly parallel to the removal of the external potential by the lens transform, while the infinite time range on the Minkowski (or SN) side is mapped to a finite time range on the global AdS (or SNH) side in both cases.


\section{The resonant approximation}

We shall now focus on the behavior of small amplitude solutions of the SNH equation.
 Since the normal modes of the linearized SNH equation (which is just the linear Schr\"odinger equation with a harmonic potential) oscillate with commensurate frequencies, nonlinear interactions can produce infinitely many resonances between the modes. In this situation, sometimes referred to as \emph{fully resonant}, the evolution of solutions with initially arbitrarily small amplitudes of order $\eps$ can lead to large nonlinear effects on timescales of order $1/\eps^2$.

An attractive position space approach to long-time evolution of the nonlinear Schr\"odinger equation with harmonic potential has been developed in \cite{Carles} (see also \cite{CMS} where this approach has been applied to the Hartree equation with harmonic potential). The key observation is that the evolution of small amplitude solutions is mostly dominated by the harmonic potential which makes the solution focus near the origin periodically in time. At the foci, the harmonic potential term is negligible and the passing through each focus can be described by a certain nonlinear scattering operator. Thus, iterations of this operator provide a stroboscopic
picture of the evolution. Unfortunately, it seems very difficult to control the iterations in the case of the SNH equation. (A similar approach to relativistic gravitational perturbations of AdS has been later attempted in \cite{position}.)

Here, to take into account the resonant interactions between linearized normal modes at leading order in $\eps$, we shall employ the \emph{resonant approximation}. We start by expanding solutions of (\ref{eqSNH}) with amplitudes of order $\eps$ in terms of the radial eigenfunctions of the $d$-dimensional isotropic harmonic oscillator:
\begin{align}
u(t,r)=\eps\,\sum_{n=0}^{\infty} \alpha_n(t) e_n(r) e^{-i E_n t},
\label{utr}
\end{align}
where the eigenvalues and normalized eigenfunctions are given by
\begin{align}
E_n=\left(2n+\frac{d}{2}\right)\omega ,\qquad e_n(r)=\omega^{d/4} \left(\frac{2n!}{\Gamma\left(n+\frac{d}{2}\right)}\right)^{1/2} L_n^{(\frac{d}{2}-1)}(\omega r^2) e^{-\omega r^2/2}\,.
\label{eqn:energy}
\end{align}
Here, $L^{(m)}_n$ denote the generalized Laguerre polynomials, which can be defined through their generating function
\beq
\sum_{n=0}^\infty t^n L^{(\alpha)}_n(x)=\frac{e^{-\frac{tx}{1-t}}}{(1-t)^{\alpha+1}}.
\eeq
Substituting \eqref{utr} into (\ref{eqSNH}), projecting on $e_n$, and using the orthogonality relation $$\int_0^\infty e_n(r) e_m(r) r^{d-1} dr=\delta_{n m},$$ we get
\begin{align}
i \dot{\alpha}_n =-\eps^2 \,\sum_{j=0}^\infty \sum_{k=0}^\infty \sum_{l=0}^\infty S_{njkl} \bar{\alpha}_j \alpha_k \alpha_l e^{i(E_n+E_j-E_k-E_l)t},
\label{eqn:eq6}
\end{align}
where we have defined
\begin{align}
S_{njkl}=\int\limits_0^\infty r dr\, \int\limits_0^\infty \,\min\{r^{d-2},s^{d-2}\}\, e_n(r)e_j(s)e_k(s)e_l(r) s\,ds \label{eqn:coeff01}.
\end{align}
We note that $\alpha_n$ vary slowly on time scales of order $1/\eps^2$, whereas the terms on the right hand side of (\ref{eqn:eq6}) with $E_n+E_j-E_k-E_l\neq 0$ oscillate much faster on time scales of order 1. It is intuitively clear that such oscillatory terms will `average out' for very small $\eps$, and this can be justified by accurate asymptotic analysis \cite{murdock, KM}. The essence of the resonant approximation (or time-averaging) is then in discarding in (\ref{eqn:eq6}) all terms with $E_n+E_j-E_k-E_l\neq 0$, and only keeping the resonant quartets satisfying $E_n+E_j-E_k-E_l=0$, which is the same as $n+j=k+l$. We furthermore redefine $-\eps^2 t$ to be the new `slow' time (whereafter $\eps$ completely disappears from the equation). The end result is a new autonomous resonant system given by
\begin{align}
i \frac{d\alpha_n}{dt} =\sum_{j=0}^\infty \sum_{k=0}^{n+j} C_{njk,n+j-k} \bar{\alpha}_j \alpha_k \alpha_{n+j-k},
\label{eqn:eq7}
\end{align}
with the interaction coefficients
\begin{align}
C_{njkl}&=\frac12\left(S_{njkl}+S_{njlk}\right).
\label{CSsymm}
\end{align}
These coefficients enjoy the index permutation symmetries
\begin{align}
C_{njkl}=C_{njlk}=C_{jnkl}=C_{klnj}. \label{eqn:symmetries}
\end{align}
It follows from \eqref{eqn:energy} and \eqref{eqn:coeff01} that
$C_{njkl}=\omega^{d/2-1} C_{njkl}|_{\omega=1}$.
From now on we will be assuming that $\omega=1$, as any other values of $\omega$ may be obtained via trivial time rescaling in \eqref{eqn:eq7}.
The integrals defining the interaction coefficients can be evaluated in a closed form using hypergeometric functions; the explicit expressions are given in Appendix~A for completeness but will not be used below.

While we are not constructing here a rigorous proof that the resonant system approximates the original equation in the appropriate long-time small-amplitude regime, and limit ourselves to a physical argument, we expect this proof to be within easy reach of the methods of modern PDE mathematics. For finite-dimensional systems of ODEs, the proof is in textbooks \cite{murdock}, while many closely related nonlinear Schr\"odinger equations have been analyzed by mathematicians, with rigorous proofs of the accuracy of the resonant approximation \cite{GHT,KM}.
(A very compact summary oriented towards the physics audience can be found in \cite{BBCE2}.)

The resonant system \eqref{eqn:eq7} is Hamiltonian
\begin{equation}\label{resH}
 i \frac{d\alpha_n}{dt}=\frac{\partial H}{\partial \bar \alpha_n}
\end{equation}
with
\begin{equation}\label{H}
 H=\frac{1}{2} \sum_{n=0}^{\infty} \sum_{j=0}^{\infty} \sum_{k=0}^{n+j} C_{njk,n+j-k} \bar\alpha_n \bar \alpha_j \alpha_k \alpha_{n+j-k}.
\end{equation}
Besides the Hamiltonian, the resonant system (\ref{eqn:eq7}) conserves the following two quantities
(corresponding to the total mass carried by the wavefunction $u$ and the total energy of the linearized theory):
\begin{equation} \label{NE}
N=\sum_{n=0}^{\infty} |\alpha_{n}|^2, \qquad
E=\sum_{n=0}^{\infty} \left(n+\frac{d}{4}\right) |\alpha_{n}|^2,
\end{equation}
which follow by Noether's theorem from the global and local phase rotation symmetries, $\alpha_n \mapsto e^{i \phi} \alpha_n$ and $\alpha_n \mapsto e^{i n \theta} \alpha_n$, respectively. For $d=4$, there are additional conservation laws, as we shall see in the next section.


\section{The resonant system in four spatial dimensions}\label{sec:4d}

As the SNH equation enjoys enhanced symmetries in four spatial dimensions, one might justifiably expect that the resonant system likewise displays special features in this case. Indeed, we shall now demonstrate that, for $d=4$, \eqref{eqn:eq7} belongs to a class of solvable cubic resonant systems introduced in \cite{Bia18}. To this end, it suffices to show that the following linear combination of the interaction coefficients
\begin{align}
\mathcal{D}_{njkl}\equiv(n+1)\tilde C_{n-1,jkl}+(j+1) \tilde C_{n,j-1,kl}-(k+1)\tilde C_{nj,k+1,l}-(l+1) \tilde C_{njk,l+1}
\label{eqn:identityd}
\end{align}
vanishes identically if $l+k=n+j-1$. Here $\tilde C_{njkl}=\frac{1}{2} (\tilde S_{njkl}+\tilde S_{njlk})$, where following \cite{Bia18} we defined $\tilde S_{njkl}=\sqrt{(n+1)(j+1)(k+1)(l+1)} S_{njkl}$. From \eqref{eqn:coeff01} and \eqref{eqn:energy}, we get for $d=4$
\begin{equation}\label{Stilde}
\tilde S_{njkl}=\int_0^\infty d\rho \int_0^\infty d\sigma \min\{\rho,\sigma\} e^{-\rho-\sigma} L_{n}^{(1)}(\rho)L_{j}^{(1)}(\sigma)L_{k}^{(1)}(\sigma)L_{l}^{(1)}(\rho).
\end{equation}
The proof that $\mathcal{D}_{njkl}\equiv 0$ if $l+k=n+j-1$, relying on identities satisfied by the Laguerre polynomials, is given in Appendix~B.

The results of \cite{Bia18} imply that any cubic resonant system of the form \eqref{eqn:eq7} with the interactions coefficients satisfying the above identity possesses an additional complex-valued conserved quantity and a three-dimensional invariant manifold on which the dynamics is completely integrable. More specifically,\footnote{In the language of \cite{Bia18}, the system \eqref{eqn:eq7} corresponds to $G=2$, where $G$ is a parameter labelling the solvable resonant systems within the large class defined in \cite{Bia18}.} the extra conserved quantity reads
\begin{align}
Z&=\sum_{n=0}^{\infty} \sqrt{(n+1)(n+2)}\, \bar{\alpha}_{n+1} \alpha_n,\label{eqn:cq4}
\end{align}
while the invariant manifold is given by the ansatz
\begin{align}
\alpha_{n}(t)=\sqrt{n+1}\,\left(b(t) + \frac{a(t)}{p(t)}\,n\right)(p(t))^{n},
\label{eqn:ansatz}
\end{align}
where $a(t),b(t)$, and $p(t)$ are complex-valued functions of time.

We shall now briefly recapitulate the main points of \cite{Bia18}, while referring the reader to the original publication for detailed derivations. The conservation of (\ref{eqn:cq4}) is shown in a fairly straightforward manner by differentiating that expression, and then applying the equations of motion \eqref{eqn:eq7} and the identity (\ref{eqn:identityd}).
Proving that the ansatz (\ref{eqn:ansatz}) is compatible with the equations of motion, on the other hand, requires some extra work.
One approach is to establish the following identities as a consequence of (\ref{eqn:identityd}):
\begin{align}
\sum_{k=0}^{n+j} \sqrt{\frac{(1+k)(1+n+j-k)}{(1+n)(1+j)}} C_{njk,n+j-k}&=1, \label{eqn:id1}\\
\sum_{k=0}^{n+j} \,k\,\sqrt{\frac{(1+k)(1+n+j-k)}{(1+n)(1+j)}} C_{njk,n+j-k} &=\frac{1}{2}(n+j), \label{eqn:id2}\\
\sum_{k=0}^{n+j}  \,k\,\sqrt{\frac{(1+k)(1+n+j-k)}{(1+n)(1+j)}} C_{njk,n+j-k} &=\frac{3}{8}(n^2+j^2)+\frac{3}{8}n j+\frac{1}{8}(n+j).\label{eqn:id3}
\end{align}
The proof, given in \cite{Bia18}, simply analyzes the condition $\mathcal{D}_{njkl}=0$ as a linear finite-difference equation for the interaction coefficients $C_{nmkl}$, and the above identities result from this analysis. While completely elementary, the proof is somewhat lengthy and we shall refrain from quoting it here in full, referring the reader to \cite{Bia18}.

Once the above identities have been established, one can substitute (\ref{eqn:ansatz})  into the equations of motion \eqref{eqn:eq7}, and the identities (\ref{eqn:id1}-\ref{eqn:id3}) will allow explicit evaluation of the sums. Both left-hand side and right-hand side of (\ref{eqn:ansatz}) descend in this manner to quadratic polynomials in $n$, and equating the coefficients of these polynomials on both sides results in a consistent set of equations for $a(t)$, $b(t)$ and $p(t)$, which can be written as
\begin{align}
i \dot{p}=&\frac{1}{16} (1+y)^2 (2 |a|^2 p (y+1)+a \bar{b}),\label{eqn:eom1b}\\
i\dot{a}=&\frac{1}{16} a (1+y)^2 (10 |a|^2 (y+1) (3 y+1)+20 (y+1)\Re(a\bar{b}\bar{p})+4 \bar{a} b p (y+1)+7|b|^2),\label{eqn:eom2b}\\
i \dot{b} =&a \bar{p}\frac{3}{8}(1+y)^4\left(2(1+2y)|a|^2+a\bar{b}\bar{p}\right)\nonumber\\
&+b(1+y)^2\left((1+y)(1+3y)|a|^2+\frac{1}{2}|b|^2+2(1+y)\Re(\bar{a}bp) \right),\label{eqn:eom3b}
\end{align}
where we have defined $y=|p|^2/(1-|p|^2)$. The existence of two conserved quantities in involution (in addition to the Hamiltonian) given by (\ref{NE}) guarantees that this reduced Hamiltonian system with three degrees of freedom is Liouville-integrable.
It is convenient to reparametrize the conserved quantities in terms of $N$, $E$, $S=\sqrt{(N^2-H)/6}$ and $Z$, which are given by
\begin{align}
N=&(1+y)^2 \left(2(1+y)(1+3y)|a|^2 +|b|^2+4 (1+y)\Re(\bar{a}bp)\right), \label{eqn:cq1b}\\
E=&(1+y)^2 \left(4 (1+y)(1+6y+6y^2)|a|^2+4(1+y)(2+3y)\Re(\bar{a}bp)+(1+2y)|b|^2\right), \label{eqn:cq2b}\\
S=&\frac{1}{2} |a|^2 (1+y)^4, \label{eqn:cq3b}\\
Z=&2(1+y)^3\left(6(1+y)(1+2y)|a|^2+6(1+y)\Re (\bar{a}bp)+|b|^2 \right) \bar{p}+2(1+y)^3 \bar{a}b. \label{eqn:cq4b}
\end{align}
Using these conservation laws, one may write down a closed equation for $p$:
\begin{align}
i \dot{p}=&\frac{1}{32(1+y)}\left(\bar{Z}-(N+E+4S)p\right).
\end{align}
One further obtains
\begin{align}
\dot{y}^2+\frac{1}{1024}\left((N-E+4S)^2+4(N^2-NE-4ES+48S^2)y+4(N^2+48S^2)y^2\right)=0,
\end{align}
and then
\begin{align}
\dot{y}^2+\omega^2 (y-y_0)^2=\mathcal{E},
\end{align}
where
\beq\nonumber
\omega=\frac{1}{16}\sqrt{N^2+48 S^2},\quad
y_0=\frac{1}{2}\left(1-\frac{E(N+4S)}{N^2+48S^2}\right),\quad
\mathcal{E}=\frac{S(4S-N)(N^2-E^2+48S^2)}{128(N^2+48S^2)}.
\eeq
The variable $y$ thus performs harmonic oscillator motions of the form
\begin{align}
y(t)=A \cos(\omega t+\phi)+y_0,
\end{align}
where
\begin{align}
A=\frac{\sqrt{2S(4S-N)(N^2-E^2+48S^2)}}{N^2+48S^2}.
\end{align}
This implies (as in a number of related physically motivated cases, as well as in the general considerations of \cite{Bia18}) that $|p|^2$ is exactly periodic in time, and by the conservation laws, $|a|^2$, $|b|^2$ and $\Re(\bar{a}bp)$ are exactly periodic in time. The same periodicity applies to the entire spectrum $|\alpha_n|^2$ for solutions within our ansatz (\ref{eqn:ansatz}). In particular, all two-mode initial data of the form $\alpha_{n\ge 2}(0)=0$ belong to our ansatz (corresponding to $p(0)=0$), and will hence display periodic exact energy returns to the initial configuration. Such initial data are often analyzed numerically in the literature on the original relativistic version of AdS perturbations. (One may, for instance, contrast the exact energy returns we have proved here for the nonrelativistic limit of AdS$_5$ with the very close but inexact energy returns  seen in the relativistic theory in AdS$_4$ \cite{returns}.) For solutions within our ansatz, as $|p|^2$ oscillates, the energy is periodically transferred to higher modes, and then it returns to the lower modes. The range of these oscillations is uniformly bounded by the considerations of \cite{Bia18}, and no turbulent behaviors emerge for solutions within the ansatz (\ref{eqn:ansatz}).

We note that no immediate analogs of the structures we displayed above (the conservation of $Z$, dynamically invariant manifolds, periodic evolution of the energy spectrum) exist in relativistic AdS gravity. At the same time, systems with relativistic self-interacting probe fields in AdS backgrounds do display such features \cite{CF,BEL}.

\begin{figure}[t]
		\centering
		\includegraphics[scale=0.38]{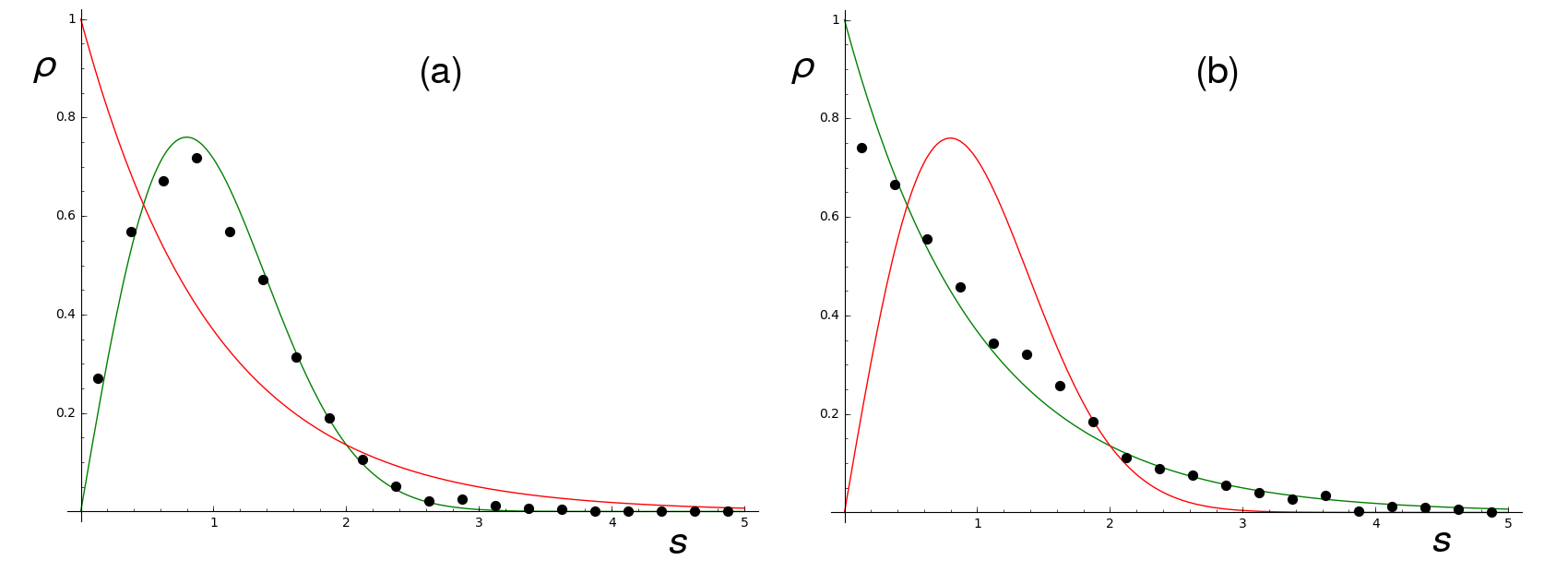}
		\caption{Eigenvalue spacing distibutions obtained for the quantized SNH resonant system in three (a) and four (b) spatial dimension for a particular eigenvalue block with $N=27$ and $E'=27$. By inspection, (a) follows the `Wigner surmise' distribution associated with chaotic systems, while (b) follows the Poisson distribution that alludes to its special structure (`partial integrability') discussed in section \ref{sec:4d}.}
		\label{fig:SNH34}
	\end{figure}

We conclude with a digression that should highlight the special character of the SNH resonant system in four spatial dimensions. Following \cite{quantres}, one can straightforwardly quantize the Hamiltonian (\ref{H}) by promoting $\alpha_n$ and $\bar\alpha_n$ to the standard creation-annihilation operators satisfying $[\hat\alpha^\dagger_n,\hat\alpha_m]=-\de_{nm}$. This `quantum resonant system' turns out to be remarkably simple and can be analyzed in great detail. The quantized versions of the conserved quantities $N$ and $E':=E-dN/4$, with $N$ and $E$ given by (\ref{NE}), have integer eigenvalues and the Hamiltonian is block-diagonal with respect to these integer eigenvalues (the matrix elements of the Hamiltonian between states with different $N$ and $E'$ vanish). Each $(N,E')$-block furthermore contains a finite number of states and can be diagonalized by ordinary matrix diagonalizations. Once the energy eigenvalues have been obtained by this diagonalization, their properties can be analyzed in any desired way. In particular, statistics of distances between the neighboring eigenvalues\footnote{At a technical level, one must properly normalize, or `unfold' these level spacings -- the details can be found in \cite{quantres}.} has been extensively investigated in the field of `quantum chaos', and is believed to be a good indicator of the integrable/chaotic features in the corresponding classical system due to the Berry-Tabor \cite{btint} and Bohigas-Giannoni-Schmit \cite{BGS} conjectures (for reviews, see \cite{GMW,haake}). More specifically, one expects that normalized level spacings $s$ obey the Poisson distribution
\beq
\rho_{Poisson}(s)=e^{-s},
\label{pois}
\eeq
for generic integrable systems (i.e., they statistically behave as distances between points randomly thrown on a line with a unit mean density), and closely follow the `Wigner surmise'
\beq
\rho_{Wigner}(s)=\frac{\pi s}2\, e^{-\pi s^2/4}.
\label{wign}
\eeq
for quantized versions of classically chaotic systems (i.e. statistically behave as distances between eigenvalues of Gaussian random matrices). We have explored this perpspective for the SNH resonant system, having picked for our statistical analysis a particular Hamiltonian block with $N=27$ and $E'=27$. The results, depicted on Fig.~\ref{fig:SNH34}, indicate that the three-dimensional resonant system closely follows the `Wigner surmise' distribution (with some statistical scatter attributed to working with a finite number of eigenvalues), while the corresponding plot for the four-dimensional system follows the Poisson distribution. This blind analysis thus suggest that no integrability features are anticipated in three dimensions (we expect similar results in more than four dimensions), while the case of four dimensions is immediately spotted as special due to a different distribution of energy eigenvalue spacings. It remains to be seen whether the Poisson curve in Fig.~\ref{fig:SNH34}b alludes to any further analytic structures for the SNH resonant system in four spatial dimensions, beyond those we have already displayed explicitly in our considerations.


\section{Higher dimensions}
We conclude with some comments on the dynamics of the SNH system and its resonant approximation in higher dimensions.
Solutions of the SN equations are known to exist at least for a finite time for initial data whose first $(d/2-2)$ spatial derivatives are square-integrable (more precisely,
the Cauchy problem for the SN equation is locally well-posed in the Sobolev space $H^s(\mathbb{R}^d)$ for $s\geq d/2-2$ \cite{Cazenave}).
For small amplitude solutions in the energy subcritical dimensions $d<6$, the norm $H^1(\mathbb{R}^d)$ is controlled by the energy and therefore conservation of energy implies global in time well-posedness in $H^1(\mathbb{R}^d)$. An analogous result has been proved for the energy subcritical SNH equation \cite{Chen}, however we are not aware of any global existence result in $d\geq 6$. In accord, the dimension $d=6$ appears critical for the resonant system \eqref{eqn:eq7} as well, which is reflected in the ultraviolet asymptotics of the interactions coefficients. Namely, for any resonant combination of nonzero indices $(n,j,k,l)$, we have for $\lambda \rightarrow \infty$
\begin{equation}\label{ultra}
C_{\lambda n, \lambda j, \lambda k, \lambda l} \sim
\lambda^{\beta}, \quad \mbox{where} \quad
\beta(d)=\begin{cases}
-1/2  \,\,\,\,\qquad  \mbox{if} \,\, d=3, \\
 (d-6)/2 \quad \mbox{if}\,\, d\geq 5.
\end{cases}
\end{equation}
In the `solvable' case $d=4$ there is a logarithmic correction to the power law: $C_{\lambda n, \lambda j, \lambda k, \lambda l} \sim \ln{\lambda}/\lambda$. (Our conclusions on the asymptotic behavior of the interaction coefficients follow from numerical studies, but it should be possible to derive this type of formulas analytically using the methods of \cite{CEV3,AT,EJ}.) As follows from \eqref{ultra}, in energy supercritical dimensions $d\ge 7$ the interaction coefficients grow with the mode number which creates hopes for interesting short wavelength behaviors in these dimensions. This matter should be investigated numerically (and, if possible, analytically), for instance, in search of turbulent behaviors of the type seen in \cite{BMR}. We leave this to future work.

\section*{Acknowledgments}

We thank Anxo Biasi and Ben Craps for discussions and collaboration on related subjects, and Garry Gibbons for a useful discussion on nonrelativistic limits. We also thank R\'emi Carles for helpful comments and references. This research has been supported by the CUniverse research promotion project (CUAASC) at Chulalongkorn University, by the Polish National Science Centre grant no.\ 2017/26/A/ST2/00530, and by the Polish Ministry of Science and Higher Education within the Diamond Grant program (grant no.\ 0143/DIA/2016/45).

\section*{Appendix A}
The integrals defining the interaction coefficients \eqref{eqn:coeff01} can be computed to yield
\begin{align}
S_{njkl}=&\frac{1}{8}A_n A_j A_k A_l \sum_{p_1=0}^j \sum_{p_2=0}^k \sum_{p_3=0}^n \sum_{p_4=0}^l a_{j,p_1} a_{k,p_2} a_{n,p_3} a_{l,p_4}\nonumber\\
&\times \left[\Gamma\left(\frac{d}{2}+p_1+p_2\right)\Gamma\left(1+p_3+p_4\right)+\Gamma\left(1+\frac{d}{2}+p_1+p_2+p_3+p_4\right)\right.\nonumber\\
&\left.\times\left(-\frac{{_2 F_1}\left(1+p_3+p_4,1+\frac{d}{2}+p_1+p_2+p_3+p_4;2+p_3+p_4;-1\right)}{1+p_3+p_4}\right.\right.\nonumber\\
&\left.\left.+\frac{{_2 F_1}\left(p_3+p_4,1+\frac{d}{2}+p_1+p_2+p_3+p_4;1+\frac{d}{2}+p_3+p_4;-1\right)}{\frac{d}{2}+p_3+p_4}\right)\right],\label{hypergeom}
\end{align}
where
\begin{equation}
A_n=\sqrt{\frac{2 n!}{\Gamma(n+d/2)}},\qquad
a_{n,p}=\frac{(-1)^{p}}{p!} \binom{n-1+d/2}{n-p}.
\end{equation}
For even $d$, the hypergeometric functions in (\ref{hypergeom}) are expressible in terms of elementary functions. In four dimensions in particular, this gives
\begin{align}
S_{njkl}=&\frac{1}{\sqrt{(n+1)(j+1)(k+1)(l+1)}}\sum_{p_1=0}^j \sum_{p_2=0}^k \sum_{p_3=0}^n \sum_{p_4=0}^l a_{j,p_1} a_{k,p_2} a_{n,p_3} a_{l,p_4}\nonumber\\
&\times \left[ (p_1+p_2+1)! (p_3+p_4)!-\frac{(p_1+p_2+p_3+p_4)!}{2^{p_1+p_2+p_3+p_4+2}}\right.\nonumber\\
&\left.+(p_1+p_2)! \sum_{p=0}^{p_1+p_2} \frac{(p+p_3+p_4)!}{p! 2^{p+p_3+p_4+2}}(p-2p_1-2p_2+p_3+p_4-1)\right].
\end{align}
One can verify by an explicit calculation that $S_{0000}=1/2$.

\section*{Appendix B}
Here we prove, using the properties of Laguerre polynomials, that for $d=4$ the quantity $\mathcal{D}_{njkl}$ defined in \eqref{eqn:identityd} vanishes identically if $l+k=n+j-1$.
We begin by proving the vanishing of $\mathcal{D}_{njkl}$  with the (unsymmetrized) coefficients $\tilde S_{njkl}$, given in \eqref{Stilde}, substituted instead of $\tilde C_{njkl}$. By definition,
\begin{align}
\mathcal{D}_{njkl}=&\int_0^\infty d\rho \int_0^\infty d\sigma \min(\rho,\sigma) e^{-\rho-\sigma}\left[(n+1)L_{n-1}^{(1)}(\rho)L_{j}^{(1)}(\sigma)L_{k}^{(1)}(\sigma)L_{l}^{(1)}(\rho)\right.\nonumber\\
&+(j+1)L_{n}^{(1)}(\rho)L_{j-1}^{(1)}(\sigma)L_{k}^{(1)}(\sigma)L_{l}^{(1)}(\rho)-(k+1)L_{n}^{(1)}(\rho)L_{j}^{(1)}(\sigma)L_{k+1}^{(1)}(\sigma)L_{l}^{(1)}(\rho)\nonumber\\
&\left.-(l+1)L_{n}^{(1)}(\rho)L_{j}^{(1)}(\sigma)L_{k}^{(1)}(\sigma)L_{l+1}^{(1)}(\rho)\right].
\end{align}
Using the identity $nL_n^{(1)}(\rho)=(n+1)L_{n+1}^{(1)}(\rho)-\rho L_{n-1}^{(2)}(\rho)$, one can write
\begin{align}
\mathcal{D}_{njkl}=&\int_0^\infty d\rho \int_0^\infty d\sigma \min(\rho,\sigma) e^{-\rho-\sigma}\left[(n+j+k-l-4)L_n^{(1)}(\rho)L_j^{(1)}(\sigma)L_k^{(1)}(\sigma)L_l^{(1)}(\rho)\right.\nonumber\\
&\left.+\rho L_{j}^{(1)}(\sigma)L_{k}^{(1)}(\sigma) \left(L_{n-1}^{(2)}(\rho)L_{l}^{(1)}(\rho)+L_{n}^{(1)}(\rho)L_{l}^{(2)}(\rho)\right)\right.\nonumber\\
&\left.+\sigma L_{n}^{(1)}(\sigma)L_{l}^{(1)}(\sigma) \left(L_{j-1}^{(2)}(\rho)L_{k}^{(1)}(\rho)+L_{j}^{(1)}(\rho)L_{k}^{(2)}(\rho)\right)\right].
\end{align}
Remembering that $l+k=n+j-1$, and applying the identities $L_n^{(1)}=L_n^{(n)}-L_{n-1}^{(2)}$ and $\partial_\rho L_n^{(1)}(\rho)=-L_{n-1}^{(2)}(\rho)$, one can simplify $\mathcal{D}_{njkl}$ further:
\begin{align}
\mathcal{D}_{njkl}=&\int_0^\infty d\rho \int_0^\infty d\sigma e^{-\rho-\sigma} \min (\rho, \sigma)(\rho+\sigma-3-\rho\partial_\rho-\sigma\partial_\sigma) L^{(1)}_n(\rho)L^{(1)}_j(\sigma)L^{(1)}_k(\sigma)L^{(1)}_l(\rho) \nonumber\\
=&\int_0^\infty d\rho \int_0^\rho d\sigma e^{-\rho-\sigma} \sigma(\rho+\sigma-3-\rho\partial_\rho-\sigma\partial_\sigma) L^{(1)}_n(\rho)L^{(1)}_j(\sigma)L^{(1)}_k(\sigma)L^{(1)}_l(\rho) \nonumber\\
&+\int_0^\infty d\rho \int_\rho^\infty d\sigma e^{-\rho-\sigma} \rho (\rho+\sigma-3-\rho\partial_\rho-\sigma\partial_\sigma) L^{(1)}_n(\rho)L^{(1)}_j(\sigma)L^{(1)}_k(\sigma)L^{(1)}_l(\rho) \nonumber\\
=&-\int_0^\infty d\rho \partial_\rho\left(\rho e^{-\rho}L^{(1)}_n(\rho)L^{(1)}_l(\rho)\right) \int_0^\rho d\sigma e^{-\sigma} \sigma L^{(1)}_j(\sigma)L^{(1)}_k(\sigma) \nonumber\\
&-\int_0^\infty d\rho e^{-\rho} L^{(1)}_n(\rho)L^{(1)}_l(\rho) \int_0^\rho d\sigma \partial_\sigma\left(\sigma^2 e^{-\sigma}L^{(1)}_j(\sigma)L^{(1)}_k(\sigma) \right)\nonumber\\
&-\int_0^\infty d\rho \partial_\rho\left(\rho^2 e^{-\rho}L^{(1)}_n(\rho)L^{(1)}_l(\rho)\right) \int_\rho^\infty d\sigma e^{-\sigma} L^{(1)}_j(\sigma)L^{(1)}_k(\sigma) \nonumber\\
&-\int_0^\infty d\rho e^{-\rho}\rho L^{(1)}_n(\rho)L^{(1)}_l(\rho) \int_\rho^\infty d\sigma \partial_\sigma\left(e^{-\sigma}\sigma L^{(1)}_j(\sigma)L^{(1)}_k(\sigma) \right).
\label{eqn:d}
\end{align}
Then the first term of the last representation may be rewritten with integration by parts
\begin{align}
&\int_0^\infty d\rho \partial_\rho\left(\rho e^{-\rho}L^{(1)}_n(\rho)L^{(1)}_l(\rho)\right) \int_0^\rho d\sigma e^{-\sigma} \sigma L^{(1)}_j(\sigma)L^{(1)}_k(\sigma) \nonumber\\
=&-\int_0^\infty d\rho e^{-2\rho} \rho^2 L_n^{(1)}(\rho)L_j^{(1)}(\rho)L_k^{(1)}(\rho)L_l^{(1)}(\rho),
\end{align}
while the second term of this sum can be simply integrated to
\begin{align}
&\int_0^\infty d\rho e^{-\rho} L^{(1)}_n(\rho)L^{(1)}_l(\rho) \int_0^\rho d\sigma \partial_\sigma\left(\sigma^2 e^{-\sigma}L^{(1)}_j(\sigma)L^{(1)}_k(\sigma) \right)\nonumber\\
=&\int_0^\infty d\rho e^{-2\rho} \rho^2 L_n^{(1)}(\rho)L_j^{(1)}(\rho)L_k^{(1)}(\rho)L_l^{(1)}(\rho).
\end{align}
The third and fourth terms are treated analogously, and the whole sum reduces to zero. We may repeat these computations for $S_{njlk}$ and also get zero. These two calculations imply that $\mathcal{D}_{njkl}$, as defined in (\ref{eqn:identityd}), is identically zero for $l+k=n+j-1$.


\end{document}